\documentclass[english,a4paper,prx,twocolumn,amsmath,amssymb,superscriptaddress]{revtex4}
\usepackage{graphicx,color,soul}

\begin{document}

\title{Crystallisation Instability in Glassforming Mixtures}

\author{Trond S. Ingebrigtsen}
\affiliation{Department of Fundamental Engineering, Institute of Industrial Science, University of Tokyo, 4-6-1 Komaba, Meguro-ku, Tokyo 153-8505, Japan.}

\author{Jeppe C. Dyre}
\affiliation{DNRF Centre ``Glass and Time'', IMFUFA, Department of Science, Systems and Models, Roskilde University, Postbox 260, DK-4000 Roskilde, Denmark.}

\author{Thomas B. Schr{\o}der}
\affiliation{DNRF Centre ``Glass and Time'', IMFUFA, Department of Science, Systems and Models, Roskilde University, Postbox 260, DK-4000 Roskilde, Denmark.}

\author{C. Patrick Royall~}
\affiliation{H.H. Wills Physics Laboratory, Tyndall Avenue, Bristol, BS8 1TL, UK.}
\affiliation{School of Chemistry, University of Bristol, Cantock's Close, Bristol, BS8 1TS, UK.}
\affiliation{Centre for Nanoscience and Quantum Information, Tyndall Avenue, Bristol, BS8 1FD, UK.}

\begin{abstract}
Crucial to gaining control over crystallisation in multicomponent materials or accurately modelling rheological behaviour of magma flows is to understand the mechanisms by which crystal nuclei form. The microscopic nature of such nuclei, however, makes this extremely hard in experiments, while computer simulations have hitherto been hampered by their short timescales and small system sizes due to limited computational power. Here we use highly-efficient GPU simulation techniques to access system sizes around 100 times larger than previous studies. This makes it possible to elucidate the nucleation mechanism in a well-studied binary glassformer. We discover that the supercooled liquid is inherently unstable for system sizes of $10,000$ particles and larger. This effect is due to compositional fluctuations leading to regions comprised of large particles only which rapidly nucleate. We argue that this mechanism provides a minimum rate of crystallisation in mixtures in general, and use our results to stabilise a model binary mixture and predict glassforming ability for the CuZr metallic glassformer.
\end{abstract}

\maketitle

\section{Introduction}
\label{sectionIntroduction}

Crystallisation in supercooled liquids has profound implications in fields as diverse as the development of amorphous materials \cite{greer}, magma flows in volcanoes \cite{cashman2000} and aqueous solutions of ions \cite{bullock2013}. Materials in question include metallic, inorganic and chalcogenide glassformers, where mixtures of a number of different elements have the effect of suppressing or controlling crystallisation \cite{berthier2011}. Alas, this tendency to crystallise hampers the exploitation of materials such as metallic glassformers as it places stringent limits on the size of the pieces that can be formed: large pieces are more likely to undergo crystal nucleation \cite{inoue2011,togashi2008}. This ``Achilles heel'' of glass formation thus limits the application of these potentially important materials, whose superior mechanical properties otherwise hold great promise  \cite{cheng2011}. It is clear that any liquid cooled below its freezing point must, for a sufficiently large system, nucleate. However, the practical limits of cooling rate versus system size required for vitrification are not known in general. In additional to these practical considerations, crystallisation is one solution to the Kauzmann paradox of vanishing configurational entropy upon which a number theories of the glass transition rest \cite{berthier2011}, thus avoiding the need to invoke any particular theoretical description of divergent viscosity in amorphous materials \cite{kauzmann1948}.

While it is known empirically that increasing the number of constituent species, introducing a size dispararity among these components together with a negative heat of mixing between the main components may serve as useful guidelines to suppress nucleation  \cite{inoue2011}, despite recent innovative approaches using model systems \cite{zhang2013jcp,toxvaerd2009} and novel sampling techniques \cite{ninarello2017,coslovich2018}, there is little fundamental understanding of the mechanisms by which glassforming mixtures crystallise. Here we consider a well-studied glassformer comprised of a binary mixture with parallelised and highly-efficient graphics card processing unit (GPU) computer simulations \cite{bailey2015} for unprecedented system sizes and simulation timescales. We reveal that such binary mixtures are inherently unstable to crystallisation due to compositional fluctuations in the supercooled liquid which lead to regions of one species which are larger than the critical nucleus size for crystallisation of that species only. We reveal similar behaviour in metallic glassformers and thus provide evidence that our results offer general insights into the design of glassformers through the use of mixtures, identifying a means to optimise their glassforming ability. These compositional fluctuations occur in the \emph{absence} of any underlying demixing behaviour driven by a thermodynamic transition. The first mixture we investigate is specifically designed \emph{not} to demix, using a non-additive attractive cross interaction between the species. Thus the fluctuations we consider are those of the mixture in meta-equilibrium (with respect to crystallisation). Therefore, the conditions we consider are distinct from enhanced crystal nucleation rates due fluctuations related to a nearby critical point \cite{tenwolde1997,vekilov2010nanoscale}.

Since its inception, the binary Lennard-Jones (LJ) model based on the metallic glassformer Nickel-Phosphorous has been a mainstay of basic systems with which to tackle the glass transition \cite{kob1994}. Prized for its simplicity, speed of computation and its stability against crystallisation, this Kob-Andersen (KA) model is among the most widely used atomistic glassformers in computer simulations. However, like the metallic glasses it was inspired by, the KA model has the potential to crystallise. It was suggested that the KA mixture might demix and freeze into a face-centred cubic (FCC) crystal of A particles and a mixed AB body-centred cubic (BCC) crystal \cite{fernandez2003pre}. Indeed recently evidence that the KA mixture would demix and partially crystallise has emerged \cite{toxvaerd2009}, consistent with the equilibrium phase diagram \cite{pedersen2018} although an analysis based on the energy landscape suggests other crystal structures might also form \cite{desouza2016}. An estimation of the nucleation barriers of the FCC and BCC crystals indicated that the barrier of the latter is much higher, suggesting that FCC might be the predominant crystal, at least from a kinetic point of view \cite{nandi2016}. However, to date very few direct simulations of crystallisation in the KA mixture have been carried out, due not least to the vast timescales involved in simulating this glassformer at sufficiently low temperatures and at large system sizes, although crystallisation in other binary mixtures has been carefully studied \cite{zhang2013jcp,jungblut2011}. The challenges with simulating the KA model under the conditions where it crystallises have left us with little idea of the mechanism, let along whether such a mechanism might apply to other materials.

Here, using optimized GPU molecular dynamics (MD) computer simulations, we address these challenges of large system sizes as well as long timescales. We reveal that the KA model possesses a fatal weakness as a glassformer: for sufficiently large system sizes, even at quite moderate supercooling (relative to experimental systems), the KA liquid is \emph{unstable} to crystallisation. That is to say, the supercooled liquid starts to crystallise on timescales comparable to the structural relaxation time. Our analysis suggests that this is an inevitable consequence of its asymmetric 4:1 composition. We show that compositional fluctuations --- rarer in the smaller ($N\lesssim1000$) systems typically simulated hitherto --- become much more frequent in the larger systems we consider here (up to $N=100,000$), and thus enable better sampling of large fluctuations in the case of bigger systems. We provide evidence that the large compositional fluctuations are exponentially distributed.

Furthermore, at the low temperatures we consider for the KA mixture ($T\sim0.40$ in reduced units), the system is at just over half the melting temperature of the one-component system $T_m=0.75$ at a comparable pressure of $P=0.67$ \cite{tenwolde}. In the one-component system (at $T \sim 0.40$) the critical nuclei are then expected to be very small. 
The aforementioned compositional fluctuations mean that regions populated only by the majority A species form which are larger than the critical nuclei. These rapidly crystallise, and once formed, we find that these nuclei grow irreversibly, albeit presumably at a slower rate than the one-component system would as the particles of the binary system must organise compositionally on the (one-component) crystal surface. Comparable behaviour of very rapid crystallisation is seen in diverse systems such as hard spheres at sufficient supercooling \cite{sanz2014,yanagishima2017}.

Our results have profound consequences for the glassforming ability of mixtures, such as metallic glasses \cite{cheng2011} and oxides \cite{salmon2013}, which we demonstrate for the case of the important material Cu$_{x}$Zr$_{1-x}$ using the compositions $x$ =  64.5, and 35.5. In CuZr, we find the same behaviour as we find in KA, underlining the generality of the crystallisation mechanism we present.

The mechanism we uncover shows that, even in the absence of a multicomponent crystal or quasicrystal with structure related to that of the liquid (which limits the glassforming ability of another binary model glassformer \cite{pedersen2010}), partial freezing into a single-component crystal is inevitable. For a given mixture, whose components are not good glassformers (as is the case with metallic glassformers and many mixtures), this places a fundamental limit on the size of the system which can be stabilised against crystallisation in the supercooled liquid regime and therefore limits the size of amorphous material which can be formed. Our results suggest that one route towards high glassforming ability is to massively increase the number of components, in order to suppress the magnitude of the regions of a single species formed by compositional fluctuations.

We make our case in the following way. First, we present our results from large-scale GPU molecular dynamics computer simulations, in which we show that the KA supercooled liquid is unstable to the formation of one-component A species crystal nuclei. To probe the consequences for the nucleation kinetics due to the fluctuations of pure A species particles, we turn to the literature on the one-component LJ system. For the larger systems that we study, we hypothesise that the number of particles in regions of pure A species formed through thermal fluctuations can exceed the critical nucleus size for the one-component A system. Now at temperature $T=0.40$, the one-component LJ liquid is indeed expected to be \emph{unstable}, in the sense that it freezes on timescales much less than the structural relaxation time \cite{yang1990,trudu2006,peng2010}. We identify crystallisation kinetics in the temperature range $0.395 \leq T \leq 0.45$ at $N=10,000$ and further investigate compositional fluctuations and crystallisation kinetics for a range of system sizes $125 \leq N \leq 100,000$ at $T = 0.40$. Remarkably, the largest compositional fluctuation increases logarithmically with system size $N$.

Finally, we explore why it is that we see crystallisation, which hitherto has only been rarely observed in the KA model. We discover that for the system sizes we consider, only the $N \gtrsim$ 1000 particle systems crystallise, while smaller systems exhibit large, reversible, fluctuations towards crystalline states and may be influenced heavily by finite-size effects. Combined with the very long duration of our GPU simulations, this may explain why crystallisation is seen only very rarely in the $N$ = 1000 particle systems typically simulated \cite{toxvaerd2009}. In fact, under an extreme value analysis, the compositional fluctuations turn out to be well described by assumptions that particles are randomly distributed. This has profound consequences: that our mechanism is general and thus it may be applied to other systems. We thus apply our analysis to other compositions of the KA model, and show that the 2:1 and 3:1 compositions, are very much more stable than the usual 4:1 mixture and that the important metallic glassformer CuZr is very stable against crystallisation (but nevertheless exhibits the same scaling behaviour as for KA).

\section{Freezing in the Kob-Andersen Model Glassformer}
\label{sectionFreezing}

\begin{figure*}
\centering
\includegraphics[width=160mm]{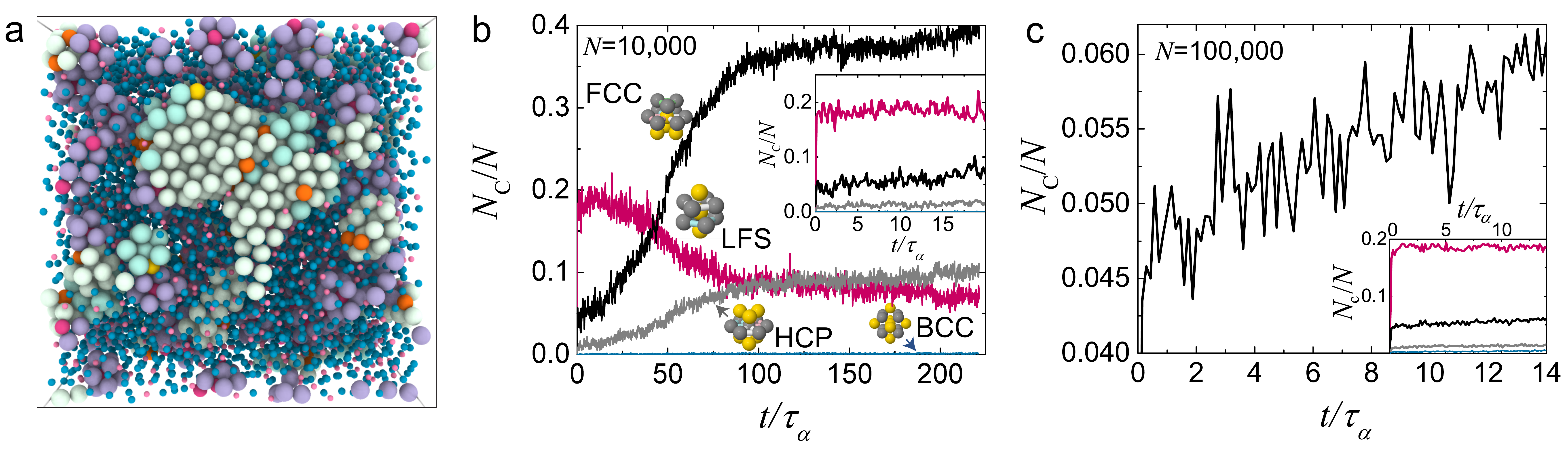}
\caption{\textbf{Particle-level structural analysis of crystallisation in the Kob-Andersen glassformer.}
\textbf{a} Particle snapshot at $t\approx 36 \tau_\alpha$ at $T =0.40$ and $N =10,000$. We observe an 
FCC crystallite of A particles in the center. Light green and yellow particles are FCC for A and B particles respectively; light blue and orange are HCP; purple and dark pink are bicapped antiprism liquid locally favoured
structure (LFS) and dark blue and light pink particles are liquid. 
\textbf{b} Population of local structures as a function of time reveals rapid crystallisation of FCC and HCP for $N=10,000$. BCC is found in small quantities and the bicapped square antiprism liquid LFS (which is incompatible with FCC and HCP) is also shown. 
Inset shows that at short times ($\lesssim 5 \tau_\alpha$), there is little growth in the crystal population. 
\textbf{c} Even more rapid crystallisation occurs when $N=100,000$, here within one relaxation time. The liquid is thus unstable. Inset: Liquid LFS and HCP populations are compared with FCC.}
\label{figTCC}
\end{figure*}

We begin the presentation of our results with a global structural analysis for $T = 0.40$ in Figs. \ref{figTCC}\textbf{b} and \textbf{c}. Additionally, Fig. \ref{figTCC}\textbf{a} shows a snapshot of a crystal nucleus comprised predominantly of the majority A species. Figures \ref{figTCC}\textbf{b} and \textbf{c} show, respectively, the time evolution of the population of liquid local structures (bicapped square antiprisms) and crystalline structures for system sizes of $N=10,000$ and $N=100,000$ (we use $N_{c}$ for the total number of particles in specific liquid or crystal structures). Here and henceforth we scale time by the structural relaxation time $\tau_{\alpha}$ which we determine as follows. First we calculate the self-part of the intermediate scattering function of the A-particles (see Methods). The resulting ``Angell'' plot we fit with the Vogel-Fulcher Tamman expression, and the fit we use to estimate $\tau_\alpha$ as shown in the supplementary material (SM). For $T=0.40$, we have that $\tau_\alpha=2.91\times10^5$ simulation time units.

We identify particles in BCC crystalline regions with bond-orientational order (BOO) parameters \cite{steinhardt1983,lechner2009,russo2012} and the remaining
structures with the topological cluster classification (TCC) algorithm \cite{malins2013fara}. A description of these order parameters is given in Methods. We see from Fig. \ref{figTCC}\textbf{b} that the liquid begins to freeze on a timescale of a few structural relaxation times $\tau_\alpha$. Thus, for these parameters of $T=0.40$ and $N=10,000$, it is hard to regard the KA mixture as anything but a very poor glassformer. We further see that the locally favoured structure (LFS) in the liquid, the bicapped square antiprism (pictured in Fig. \ref{figTCC}\textbf{b}), transforms into the crystal in much the same way as in one-component hard spheres where again the liquid LFS
competes with the crystal symmetry \cite{taffs2013}. Here of course we have a binary system, but the predominant crystal structures we find are FCC and hexagonal close-packed (HCP) of the  large A species only, and very little BCC; we observe no growth of a mixed AB crystal on the timescale of the simulation. The former observation is in keeping with predictions that the crystal nucleation barrier is very much higher in the case of BCC \cite{nandi2016}. Henceforth we neglect the BCC structure and focus on the HCP and FCC crystals. In Fig. \ref{figTCC}\textbf{b}, we see that there seems to be very little incubation time \emph{i.e.} the KA model glassformer appears \emph{unstable} to crystallisation. However close inspection (Fig. \ref{figTCC}\textbf{b} inset) reveals that for timescales of a few $\tau_\alpha$, the fluctuations in crystal population are larger than the increase, so the liquid may in fact be regarded as metastable on short timescales. In Fig. \ref{figTCC}\textbf{c}, we show that upon a further increase of system size, to $N=100,000$, this short time metastability vanishes; even on timescales comparable to the structural relaxation time the number of particles in FCC environments grows irreversibly.

\begin{figure*}
\centering
\includegraphics[width=120mm]{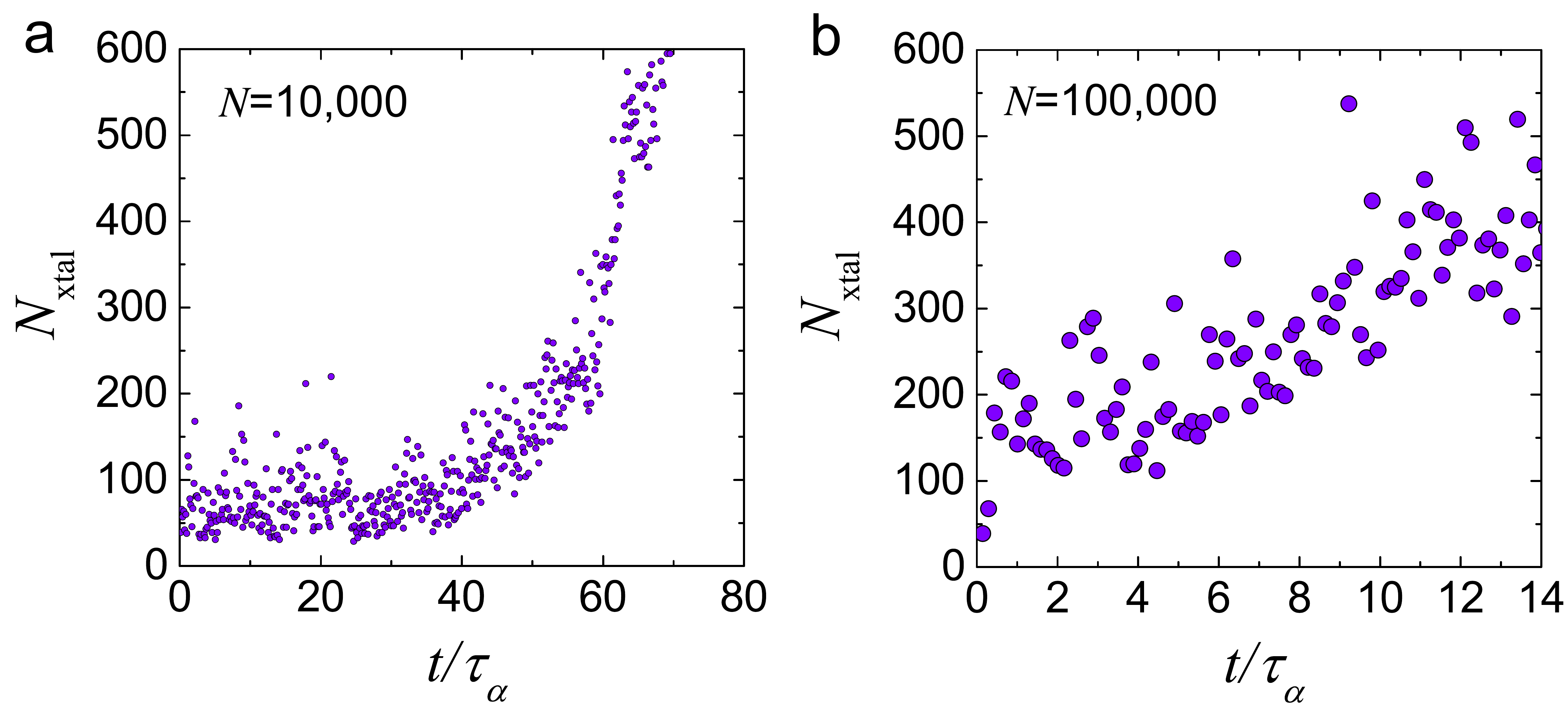}
\caption{\textbf{Time-evolution of the largest crystalline region.}
 All data are for temperature $T=0.40$. 
\textbf{a} Run with incubation period of around 40$\tau_\alpha$ prior to growth of crystalline region ($N=10,000$).
\textbf{b} Immediate crystal growth at $N=100,000$. }
\label{figLargestXtal}
\end{figure*}

We now consider the formation of critical crystal nuclei and estimate their size. In Figs. \ref{figLargestXtal}\textbf{a} and \textbf{b}, we show the number of particles $N_\mathrm{xtal}$ in the largest connected region of crystal particles (HCP or FCC), determined by requiring that they are connected through the faces between Voronoinpolyhedra surrounding each particle (see also TCC section in Methods). All our runs for $N=10,000$ showed a similar behaviour, with some change in the incubation time prior to nucleation of a crystal large enough to grow. Here we select a run with a relatively long incubation period (Fig. \ref{figLargestXtal}\textbf{a}). We see that the crystal nuclei are smaller than around 100 particles for around $40\tau_\alpha$ before growing. These data enable us to infer a critical nucleus size of approximately 100 particles for $T=0.40$ and $N=10,000$. Figure \ref{figLargestXtal}\textbf{b} shows the run at $N$ = 100,000 where crystal growth is immediate and thus it is difficult to infer a critical nucleus size with this time resolution. In other words, the KA model glassformer is \emph{unstable} to crystallisation. That is to say, on timescales of the order of the relaxation time, the size of the largest crystal region does not fluctuate in a manner suggestive of equilibrium or a long-lived metastable state, but exhibits signs of irreversible increase of the size of crystal nuclei.

\begin{figure*}
\centering
\includegraphics[width=120mm]{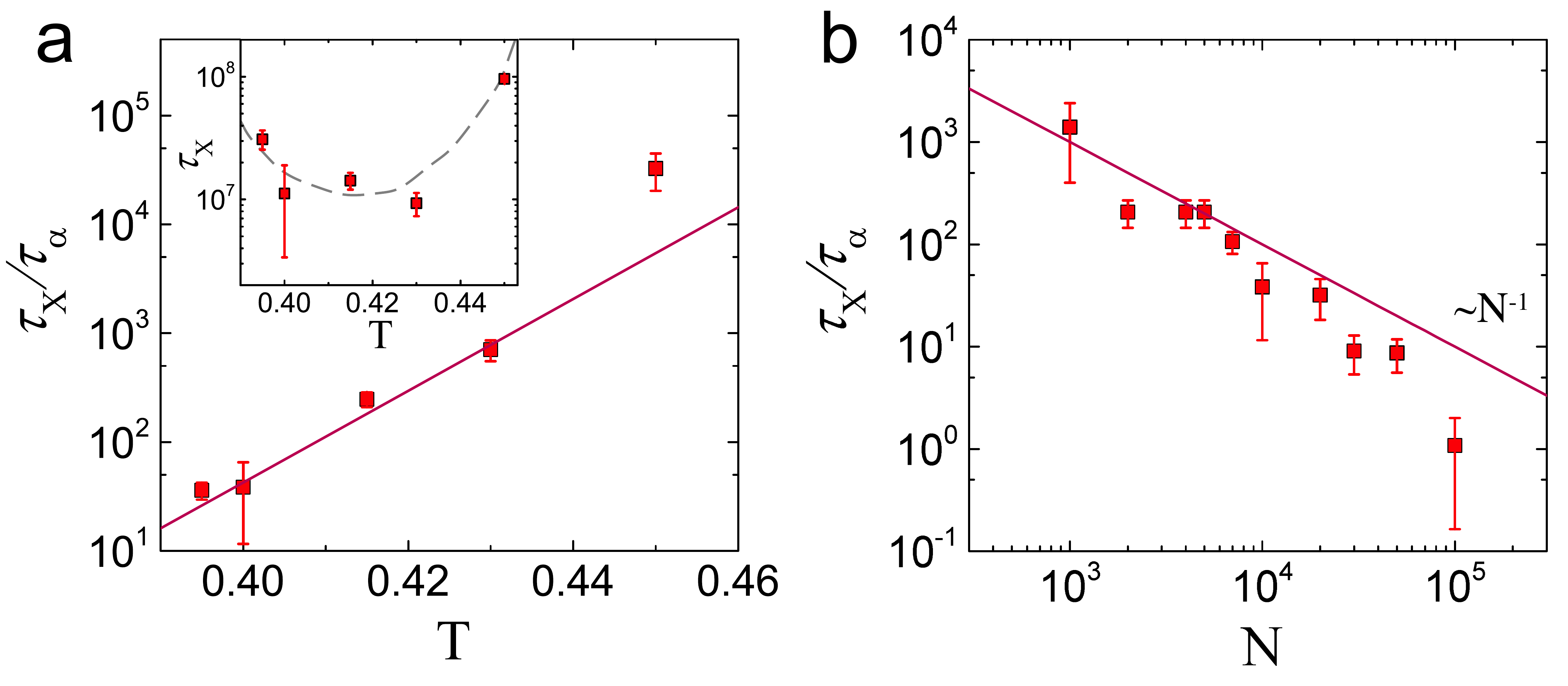}
\caption{\textbf{Crystallisation times $\tau_X$ with respect to temperature and system size.}
\textbf{a} Crystallisation as a function of temperature $T$ at $N =10,000$. Line is a fit to $\tau_X/\tau_{\alpha} \sim \mathrm{e}^{AT}$.
Inset: Crystallisation times $\tau_X$ along with a curve to guide the eye (dashed line).
\textbf{b} Crystallisation as a function of system size at $T=0.40$. Line represents expected $1/N$ scaling for $\tau_X$ in the case of a constant nucleation rate. Faster than constant nucleation rate per unit volume is noted at large system sizes. Methods of determination of $\tau_X$ are described in the text.}
\label{figNucleation}
\end{figure*}

Next, we consider the statistics of the crystallisation in the KA glassformer. From the ten runs we performed for $N = 10,000$ (all of which crystallised), we determine the mean nucleation time from $\tau_{X}=\sum_{i=1}^{n} t_{X(i)} / n$, where $n$ is the total number of simulations, to be $\tau_{X}= 38.4\pm26.8 \tau_\alpha$ ($\tau_\alpha=2.91\times10^5$ simulation time units, and the error is the standard deviation). Here $t_{X(i)}$ is the time when the size of the largest crystal region reaches, and does not drop below, 100 particles. At higher temperatures, the driving force for crystallisation is of course reduced, but the dynamics is much faster. We find that the system does crystallise at higher temperatures (we probed up to $T$ = 0.45) but that not all the runs do so. In this case, we determine the mean  nucleation time for each state point following the method of Ref. \cite{filion2011}. In particular we presume that nucleation is exponentially distributed in time, such that the probability of a nucleation event happening at time $t$ is $p(t)=1/\tau_{X}\exp(-t/\tau_X)$. The probability that a given run of length $t_\mathrm{run}$ crystallises is then $\int_0^{t_\mathrm{run}} p(t) dt=1 -\exp(-t_\mathrm{run}/\tau_X)$. The fraction of runs which crystallised then gives us $\tau_X$. Errors are estimated by considering the case that one more, or one fewer, simulation runs underwent crystallisation.

We see from Fig. \ref{figNucleation}\textbf{a} that, when scaled by the relaxation time, the time to crystallise drops rapidly with temperature at $N = 10,000$ and that well before the dynamical divergence temperature predicted from a Vogel-Fulcher-Tamman fit to the temperature dependence of the structural relaxation time  $(T_0\approx0.30$, see SM), the crystallisation time $\tau_{X}$ is expected to fall below $\tau_\alpha$ at $T\approx 0.38$. Of course this observation rests on only four data points, but given the significant magnitude of the fall in $\tau_X/\tau_{\alpha}$ with temperature, we are confident that, were this trend to continue, the observation that for some $T>T_0$, $\tau_X<\tau_\alpha$ would hold. 
In the range $T\lesssim 0.43$, we find an exponential scaling with temperature, $\tau_X/\tau_\alpha \sim\mathrm{e}^{AT}$ with $A \approx 97.$ We pursue this scaling because $\tau_\alpha$ is the principle timescale in the supercooled liquid.

When we simply plot the crystallisation time in simulation time units (Fig. \ref{figNucleation}\textbf{a} inset), we make two observations. Firstly, the absolute crystallisation time does not change hugely (around one order of magnitude) throughout the temperature range in question, while the relaxation time changes by three orders of magnitude. Secondly, there is an upturn at the lowest temperature that we consider, $T=0.395$. However, this is dwarfed by the increase in relaxation time, so the scaled quantity $\tau_X/\tau_\alpha$ continues to drop.

Turning to the system size dependence of nucleation in Fig. \ref{figNucleation}\textbf{b} we find, as expected, that the time to crystallise drops with system size. Now for a constant nucleation rate, one expects a system size scaling of $\tau_X \sim 1/N$. This is notable, given that finite size effects in small systems and artefacts related to periodic images have long been known to have a major influence in nucleation rates determined from computer simulation \cite{honeycutt1986,streitz2006}.

This form is plotted as the straight line in Fig. \ref{figNucleation}\textbf{b}. Now our larger system sizes $N=10,000$ and $100,000$ do not seem to follow this scaling but crystallise more rapidly than this scaling predicts. We note that for our smallest system which crystallised, $N=1000$, only one run crystallised, so we place little confidence in the crystallisation time we estimate. However, our data are consistent with previous work, also at the same system size, in which two runs crystallised \cite{toxvaerd2009}. This may suggest some change in the mechanism of crystallisation with increasing $N$, however, as we shall see below, the mechanism we propose naturally incorporates a system size dependence where the crystallisation time should drop faster than $1/N$. However, we emphasise that even larger system sizes would be desirable to elucidate this point further, firstly because one imagines that the $1/N$ scaling should ultimately be recovered \cite{streitz2006} and secondly because, despite the $1/N$ scaling observed in the \emph{smaller} systems that we consider, finite size effects in this size regime certainly cannot be ruled out.

\section{Composition Fluctuations}
\label{sectionComposition}

\begin{figure*}
\centering
\includegraphics[width=175mm]{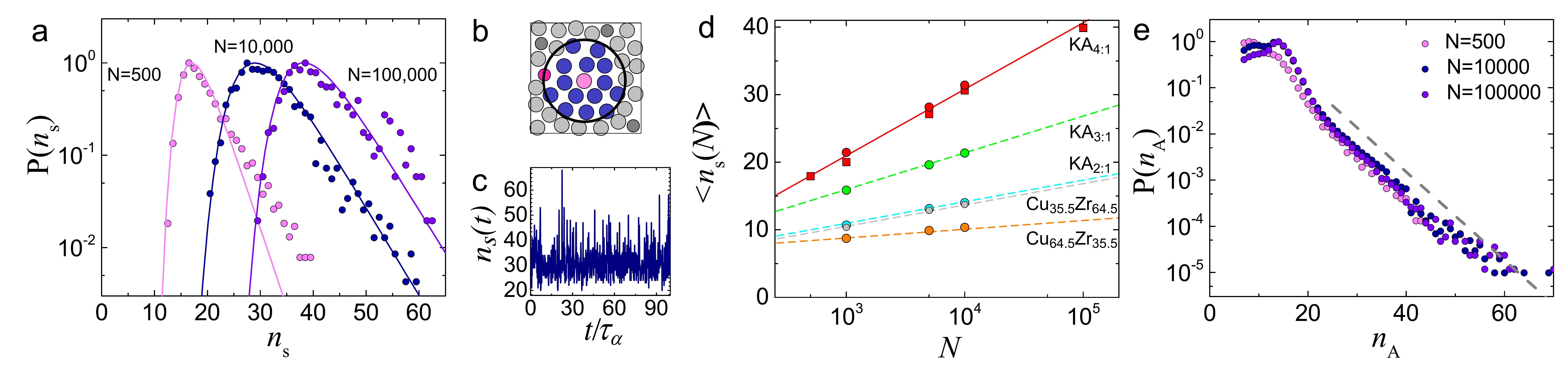}
\caption{\textbf{Compositional fluctuations of majority A-particles.} 
\textbf{a} Distributions of \textit{largest} compositional fluctuations of A-particles $n_s$ for several system sizes $N$ at $T$ = 0.40. Each system size is fitted with a Gumbel distribution (see text). 
\textbf{b} Schematic indicating the order parameter $n_s$ for
compositional fluctuations. Central pink particle is an A species under consideration and dark pink particle its nearest B species. Blue particles are A particles lying within the circle as shown. The compositional fluctuation shown has 15 A particles.
\textbf{c} Time-evolution of $n_{s}$ for the run which remained liquid the longest (for around $100 \tau_\alpha$). 
\textbf{d} Scaling of the median $\langle n_{s} \rangle$ calculated from fitted Gumbel distributions with system size; lines are fits (see text). Shown is data for the KA mixture at various compositions along with Cu$_{x}$Z$r_{1-x}$ metallic glassformers (see Methods section for details). Circles are constant pressure data ($P$ = 0) and squares are constant density data ($\rho$ = 1.204). For the KA mixture constant pressure data is taken in the NVT ensemble fixing the mean pressure at $\langle P \rangle = 0$ and $T$ = 0.80 and constant density data is taken at $T$ = 0.40. $Cu_{x}Zr_{1-x}$ is simulated in the NPT ensemble at $T$ = 1270K. \textbf{e} Distribution for all compositional fluctuations for several system sizes (4:1 KA mixture) at $T$ = 0.40. The distribution is independent of system size above $n_{A} > 25$. Dashed line denotes exponential decay with decay constant $\lambda = 0.22$ (see text).}
\label{figJeppeLength}
\end{figure*}

We showed in the previous section that the KA mixture crystallizes into an A species dominated FCC  crystal. This process is facilitated by the occurrence of compositional fluctuations in the mixture (see Fig. \ref{figJeppeLength}). Consistent with these results we find also that the liquid B particles themselves do not form extended regions (not shown). Our results also pose a rather unexpected question: given its short nucleation time (in terms of the relaxation time), why then is the KA model a glassformer at all? We believe some of the answer lies with the system-size dependence of the compositional fluctuations, combined with the fact that our simulations are able to probe metastable liquids at lower temperatures and for longer times than previously.

Let us suppose, as we concluded from Fig. \ref{figLargestXtal}, that we need a fluctuation of $n_A=100$ A particles in a region entirely devoid of B particles to initiate crystallisation. Now in the 4:1 KA mixture, the minimum system size for such a fluctuation is $N=125$. However, we expect the likelihood of a fluctuation of 100 A particles in such a system to be vanishingly small. Nevertheless, we would like to know the scaling of the likelihood of a fluctuation of $n_A=100$ with $N$. We expect that for sufficiently large system sizes, the probability of finding an $n_A=100$ fluctuation is linear in the system size. However, this argument presumes that the fluctuations are very much smaller than the system size. Clearly, for $N=125$, they are not. The question is, how big must the system be before the fluctuation distribution is independent of the system size?

To address this question, we consider now the relationship between system size and the A-particle fluctuations at $T = 0.40$. To quantify the fluctuations, we use the measure illustrated in Fig. \ref{figJeppeLength}\textbf{b}. Here we seek to find the largest region that is occupied purely by A particles. To estimate this quantity, we identify the largest sphere, centred on an A particle which contains only A particles. We assume that the sphere which contains the largest number of A particles is the largest sphere. Thus we determine the distance to the nearest B particle for all A particles. We then count the number of (A) particles in a sphere centered on each A particle starting from the largest distance between A and B particles and iteratively avoiding particles already contained in a previous sphere. To avoid the effects of crystal nuclei of A particles, we accept a maximum of 10\% of the particles in the sphere to be in a crystalline environment. After the iteration has terminated we select the largest fluctuation of A particles. This measure we term $n_s$ and its time evolution is shown in Fig. \ref{figJeppeLength}\textbf{c}. Henceforth we only consider data where the system has yet to crystallise, under our criterion of a nucleus size of less than 100 particles. We emphasise that this quantity is just one measure of the compositional fluctuations in the mixture and others are possible (see also SM \cite{SM}).  Furthermore, simply because the system has not yet crystallised does not mean that its properties are stationary, as shown in Figs. \ref{figTCC} and \ref{figLargestXtal}. However, it is still instructive to apply the same metric for the larger systems as for the smaller systems (which are stationary for a times beyond the structural relaxation time), and this we do, with the caveat that the distributions are sampled from a non-stationary system.

In Fig. \ref{figJeppeLength}\textbf{a} we see that, for $N=500$, $10,000$ and $100,000$, the distribution of $n_{s}$ has a significant dependence on the system size $N$. Two effects are apparent. Firstly, the larger system has \emph{on average} larger fluctuations of regions of A particles. Second, the distribution has a ``fat tail'' indicating larger  fluctuations than a symmetric distribution such as a Gaussian would predict. These results are consistent with the faster than expected decrease in crystallisation time observed in Fig. \ref{figNucleation}\textbf{b} and we conjecture this $N$-dependence of the largest compositional fluctuation to be its origin. However, are the observed fluctuations in Fig. \ref{figJeppeLength}\textbf{a} sufficient for crystallisation at $T=0.40$? From Fig. \ref{figJeppeLength}\textbf{a}, we see that if the critical nucleus of A particles requires more than 100 particles, then composition fluctuations capable of generating such a nucleus are more likely to be found in larger systems.

What can we say about the origin of the distribution of thecompositional fluctuations in Fig. \ref{figJeppeLength}\textbf{a}? Let us suppose that the distribution of all compositional fluctuations in the system is exponentially distributed $P(n_{A}) \propto \exp(-n_{A} \lambda)$ where $n_{A}$ is the number of $A$ particles around a given $A$ particle that are closer than the nearest $B$ particle and $\lambda$ is the decay constant. The extreme values of such a distribution, \emph{i.e.} those fluctuations large enough to initiate nucleation, should then follow a Gumbel distribution given by $P(n_{s}) \propto e^{-(z + e^{-z})}$ in which $ z \equiv (n_s - \mu)/\beta$ where $\mu$ is the mode of the probability distribution, \emph{i.e.} the highest probability point and $\beta$ is the scale of the function. Furthermore, the median of the extremes of an exponentially distributed process follows $m = 1/\lambda \, [\ln n - \ln(\ln(2))]$ in which $n$ is the number of samples \cite{bovier}.

In Fig. \ref{figJeppeLength}\textbf{e}, for several system sizes we plot the distribution of \emph{all} A particle fluctuations $P(n_{A})$ (rather than just the largest). The dashed line indicates that, for large $n_{A}$, $P(n_{A})$ exhibits an exponential decay virtually independent of the system size. Motivated by this fact we fit the Gumbel distribution to $P(n_s)$ in Fig. \ref{figJeppeLength}\textbf{a} (full lines). For $N\gtrsim 10,000$, the agreement is remarkable. Moreover, the median of the fitted Gumbel distributions $\langle n_s \rangle$ exhibits a logarithmic dependence on the system size $N$ as shown in Fig. \ref{figJeppeLength}\textbf{d}. We thus conclude that because the scaling follows the Gumbel distribution, the largest compositional fluctuation is consistent with exponentially distributed fluctuations. In fact, from Fig. \ref{figJeppeLength}\textbf{e} we find that the decay constant of the exponential is $\lambda \approx 0.22$. This value of $\lambda$ corresponds to a completely random distribution of $A$ and $B$ particles indicating that the large clusters are mainly entropic and thus present irrespective of the particular system;  the probability to find $n$ A-particles in a cluster of $n$ particles, assuming indistinguishable A and B particles, is $P(n) = (x_{A})^{n} = \exp(n\cdot\ln(0.8)) \approx \exp(-0.22 \cdot n)$.

Merely demonstrating the existence and size of these fluctuations is not sufficient. We need to show that they are sufficiently long-lived to initiate the crystallisation as well. In order to answer this question, we now consider dynamics. At $T=0.40$, as noted above, the structural relaxation time $\tau_\alpha=2.91\times10^5$. This is wildly in excess of the nucleation time in the one-component system at these temperatures \cite{yang1990,peng2010}. Thus if such compositional fluctuations persist on the timescales comparable to that of structural relaxation in the supercooled liquid $\tau_\alpha$, we have that there is \emph{little reason to suppose that the KA model would be a good glassformer}.

Another possibility for the crystallisation mechanism is enhancement of nucleation related to density fluctuations. Now the liquid-gas binodal has been measured as lying at a  temperature not much less than the $T \sim 0.40$ to which we simulate \cite{testard2014}. It is conceivable that some density fluctuations related to the proximity of liquid-gas phase separation might act to enhance nucleation, as is known for protein-like systems \cite{tenwolde1997,vekilov2010nanoscale}. However the system is not in the two-phase region and moreover the density of $\rho$ = 1.204 we consider is much higher than the critical isochore (around 0.3) \cite{testard2014}. In any case, any such nucleation enhancement would still need to invoke a mechanism for A-B demixing, which is what we provide above. Indeed to observe demixing in similar binary systems, one needs to weaken the interaction between the species so that it is again non-additive but \emph{weaker} than the additive case, \emph{i.e.} a positive enthalpy of mixing \cite{roy2016}. In short, the \emph{compositional} fluctuations we identify here are unrelated to the density fluctuations known to enhance nucleation in (effective one-component) protein-like systems \cite{tenwolde1997,vekilov2010nanoscale}.

We now consider the consequences of our choice of an instantaneous quench protocol (see Methods). In Supplementary Fig. S3, we show that the median of the largest region of liquid A particles, $\langle n_s(T) \rangle$, shows very little dependence upon temperature.  Thus, as the system samples from a nearly temperature independent distribution and due to the long mean nucleation time for $T  > 0.40$ (more than 100$\tau_\alpha$), we argue that our quenching protocol does not affect our conclusions to any significant extent.

\section{Towards a Minimum Crystallisation Rate in Mixtures}
\label{sectionTowards}

The good agreement with the Gumbel distribution in Fig. \ref{figJeppeLength}\textbf{a} has two important consequences. Firstly it suggests that the mechanism we identify should be fairly independent of the system under consideration and second, that the fluctuations $\langle n_s\rangle$ are, as we have seen, largely independent of temperature. Before investigating other materials, we consider the temperature independence which suggests that the scaling leading to large compositional fluctuations may be identified at high temperature, without recourse to simulations of the deeply supercooled liquid. This suggests that it may be possible to use our approach to \emph{predict} the glassforming ability of 
mixtures in the liquid state. To explore this idea, we consider again the KA mixture. Usually, as above, the 4:1 system is simulated, but upon changing the composition to be more equimolar, we expect smaller regions of pure A particles. We focus on the 3:1 and 2:1 KA mixtures at zero pressure and at the higher temperature of $T = 0.80$, where the relaxation times of these compositions  are comparable \cite{crowther2015}. In Fig. \ref{figJeppeLength}\textbf{d} we see that at zero pressure at the higher temperature, the 4:1 mixture has a value of $\langle n_s \rangle$ very similar to that at which we see crystallisation at $T = 0.40$. We infer that the change in pressure has little effect upon the compositional fluctuations, which is reasonable as they are largely random, according to the exponential distribution (Fig. \ref{figJeppeLength}\textbf{e}).

As expected, the 3:1 and 2:1 mixtures in Fig. \ref{figJeppeLength}\textbf{c} have very much smaller values of $\langle n_s \rangle$, as they are closer to being equimolar. To predict where these compositions might crystallise we fit each composition to a logarithmic increase as indicated by the dashed lines in Fig. \ref{figJeppeLength}\textbf{c}. We are thus able to predict the size at which the 3:1 and 2:1 KA systems reach the value of $\langle n_s(N) \rangle= 31$ (corresponding to the 4:1 system with $N\approx10,000$). The sizes we find are $N=5.8\times10^5$ and $1.2\times10^9$ for the 3:1 and 2:1 compositions respectively. Thus we expect that, for the mechanism of crystallisation we consider here, the 2:1 and 3:1 compositions should be very much better glassformers than the usual 4:1 system. In fact, we confirm this by very lengthy simulations of the 3:1 and 2:1 KA systems at comparable supercoolings and $N$ = 10,000, and 100,000 where we were unable to crystallise the systems.

Note that we are only considering the crystallisation mechanism based on compositional fluctuations. While we expect that the mechanism we present here will be present in mixtures crystallisation may be dominated by other, faster, mechanisms. For example the 1:1 KA mixture forms a BCC crystal quite rapidly \cite{fernandez2003pre}. However we emphasise that the mechanism we propose is expected to be \emph{generally} present in mixtures.  And since the compositional fluctuations exhibit so little dependence on temperature, within the range studied, it should be possible to predict the glassforming ability of, for example, metallic glassformers using simulations at liquid temperatures which are reasonably straightforward to carry out, owing to the short relaxation time. Therefore our results may be used as the basis of an optimisation strategy for producing metallic glasses which show high glassforming ability.

We now consider the predictive power of our methods for other materials. In particular the metallic glassformer Cu$_{64.5}$Zr$_{35.5}$ using Embedded Atom Model (EAM) simulations (see Methods for more details). In the SM we show that, like the KA mixture, the extreme values of the composition fluctuations in CuZr also follow a Gumbel distribution. The system size dependence of $\langle n_s(N) \rangle$, where we consider fluctuations of the majority species is shown in Fig. \ref{figJeppeLength}\textbf{d}. Again we see the logarithmic scaling, moreover CuZr has  weaker fluctuations compared to the KA model. Making the significant assumption that $\langle n_s(N) \rangle= 31$ corresponds to fluctuations sufficient to form stable crystal nuclei of Cu, we find that Cu$_{64.5}$Zr$_{35.5}$ should be stable up to system sizes of $N=3.5\times10^{16}$. For the Cu$_{35.5}$Zr$_{64.5}$ mixture we obtain $N=1.0\times10^9$. We reiterate that our method provides a minimum crystallisation mechanism, and, like the KA model, Cu$_{x}$Zr$_{1-x}$ can exhibit other routes to crystallisation \cite{ryltsev2016}. Moreover one should check that the pure Cu or Zr regions of 100 atoms do crystallise (like the one-component A regions in the Kob-Andersen model, Fig. \ref{figLargestXtal}), though given that these elements do readily crystallise, this can only be a quantitative consideration with no bearing on the qualitative mechanism we propose. Furthermore the change in magnitude of the fluctuations between the symmetric CuZr mixtures such that the Cu-rich mixture exhibits smaller fluctuations is intriguing, given that this composition is close to the eutectic point of the CuZr mixture \cite{tang2012}. Furthermore, the mechanism we have identified in no sense leads always to the equilibrium crystal. The phase diagram of the KA mixture has recently been determined, and shows a number of binary crystals, which do not form by this route \cite{pedersen2018}. The same holds for CuZr  \cite{tang2012}.

Finally we consider the consequences for the long-term stability of supercooled mixtures. We have shown that growing crystal nuclei are expected in mixtures in general. But by how much should they grow? By considering the KA mixture, the growth of FCC nuclei of A particles will deplete the remaining liquid of A particles. This will tend to slow and may even arrest the growth of the one-component A crystals. In the case of the KA system, we note that if the liquid approaches a 1:1 composition then crystallisation, not of the one-component FCC, but of the 1:1 composition BCC crystal may be expected. Given the small dimensions of the nuclei we find, and despite the developments we present here, our simulations are still small compared to experimental system sizes, and thus it seems reasonable to suppose that the final material may be comprised of \emph{nano-crystals}. Nanocrystals are known to have important consequences for the mechanical properties of glassforming materials \cite{inoue2005}. While this behaviour has been seen in experiments \cite{sahu2010}, our work suggests that such nanocrystals may be rather prevalent in metallic glasses. Because identifying tiny crystalline regions is hard with x-ray scattering, rather requiring techniques such as fluctuation TEM or 3D atom probe tomography \cite{sahu2010}, it is possible that such nanocrystals may go undetected.

\section{Discussion and Conclusions}
\label{sectionDiscussion}

We have identified a new mechanism of crystallisation in multicompoent systems. We begin by carrying out large scale simulations of a widely used model glassformer. Our results show that this system possesses a fatal flaw as a glassformer which is general to mixtures: local compositional fluctuations lead to regions populated only by the majority species. These regions of one species can be larger than the critical crystal nucleus size of the one-component system under similar conditions. Nucleation in these pure A regions is fast on the timescale of this deeply supercooled liquid, apparently requiring little re-arrangement of the particles, as is known to be the case for hard spheres at deep supercooling \cite{sanz2014,yanagishima2017}. The reason for the latter might also be that the A-particle rich regions can self-organize due to high density into favorable local structures for crystallization as seen for single-component systems \cite{russo2012}.

For the system sizes we consider, fluctuations of pure A regions grow significantly with system size. Simultaneously, we find evidence that our largest systems crystallise faster than expected from smaller systems in the case of a constant nucleation rate per unit volume. Our findings are important for two reasons: firstly, these results explain why earlier simulations with fewer particles and shorter run-times did not find crystallisation. Secondly, the results we reveal here pose a fundamental challenge for the development of glassforming materials: mixtures whose components crystallise easily are themselves inherently unstable to crystallisation and thus ultimately compromised as glassformers.

While the binary model we have considered forms a rather basic system, it nonetheless exhibits the use of a mixture to suppress crystallisation, as is typically employed in metallic and inorganic glassformers and is encountered in vitreous magmas. We expect the mechanism we have uncovered to be general in mixtures: although prevalent and accessible to computer simulation for the 4:1 binary system we consider, we expect the same mechanism will operate for more general binary mixtures, and indeed for multicomponent systems frequently employed in the quest for ever-better glassforming alloys \cite{cheng2011}. We demonstrate this by considering the well-studied CuZr metallic glassformer. Evidence in support of the mechanism we find has been seen in some metallic glasses \cite{sahu2010} and we suggest that the presence of such nanocrystals as we identify here would be worth investigating in metallic glasses.

That the compositional fluctuations are rather random and insensitive to temperature, suggests that simulations in the liquid where the dynamics are much faster may be used to predict the system size at which crystallisation may be expected. We have demonstrated this principle using the 3:1 and 2:1 KA mixtures and have predicted that both can reach system sizes, for comparable simulation times and supercoolings, very much larger than the usual 4:1 mixture before crystallisation occurs. Building on these ideas we probed two compositions of the key metallic glassformer CuZr. Compositional fluctuations, though somewhat weaker than in the KA model system, are nevertheless present and follow the same logarithmic scaling. We therefore propose that compositional fluctuations present a general mechanism for crystallisation. In addition to the metallic glasses we consider here, an intriguing case is aqueous ionic solutions. Here crystallisation of water occurs through segregation to ion-rich and ion-poor regions, the latter being where the ice nucleates, which appears similar to that we observe here, for the A particles \cite{bullock2013}. However, the various anomalies in the thermodynamic behaviour water of water, not least increasing fluctuations, which may be related to an (avoided) liquid-liquid transition \cite{moore2009,palmer2014}, mean that further study of that system would be needed to ascertain whether the mechanism we have identified here dominates water crystallisation in some aqueous solutions.

Crystallisation via segregation thus forms a lower bound to nucleation: other mechanisms involving more complex crystal structures may prove faster, as indeed seems to be the case for some models \cite{pedersen2010,ryltsev2016} and for certain compositions of the Kob-Andersen \cite{fernandez2003pre} and CuZr \cite{ryltsev2016} models considered here. Nevertheless we have shown that liquids which rely on mixing for their stability against crystallisation are fundamentally compromised and provide a principle by which their glassforming ability may be optimised.

\subsection*{Acknowledgements}
It is a pleasure to thank Nick Bailey, Daniele Coslovich, Daan Frenkel, Felix H\"{o}fling, Peter Harrowell, Takeshi Kawasaki, Rob Jack, Ken Kelton, Heidy Madre, Kunimasa Miyazaki, John Russo, Hajime Tanaka, and Jianguo Wang for stimulating discussions. CPR acknowledges the Royal Society, and Kyoto University SPIRITS fund. European Research Council (ERC consolidator grant NANOPRS, project number 617266) for financial support. The authors are grateful to Ioatzin Rios de Anda for inspirational help with the figures.

\section*{Methods}
\noindent
\textbf{Simulation and model details. --- } 
We simulate the KA binary mixture in the NVT ensemble (Nose-Hoover dynamics) at $\rho = 1.204$ using the Roskilde University Molecular Dynamics (RUMD) package \cite{bailey2015} optimized for highly-efficient GPU simulations; the longest simulations took more than 100 days. The interatomic interactions of
the 4:1 binary mixture  is defined by $v_{ij}(r) = \epsilon_{\alpha   \beta}\big[(\frac{\sigma_{\alpha \beta}}{r})^{12} - (\frac{\sigma_{\alpha \beta}}{r})^{6}]$ with parameters $\sigma_{AB} = 0.80$, $\sigma_{BB} = 0.88$ and $\epsilon_{AB} = 1.50$, $\epsilon_{BB} = 0.50$ ($\alpha$, $\beta$ = A, B). We employ a unit system in which
$\sigma_{AA}$ = 1, $\epsilon_{AA}$ = 1, and $m_{A} = m_{B}$ = 1. We study system sizes in the range $N$ = 125, 250, 500, 1000, 5000, 10,000, 100,000 at $T=0.40$ and several different temperatures $T$ = 0.395, 0.40, 0.415, 0.43, 0.45 at $N$ = 10,000. The protocol for studying crystallisation in the KA mixture is identical for all temperatures and system sizes studied. We equilibrate at $T$ = 2.00 and then perform an instantaneous quench to low temperatures simulating between 9 and 36 billion time steps after the quench ($\Delta t$ = 0.0025). For each temperature and system size we perform10 independent quenches. Additionally, 4:1, 3:1 and 2:1 KA mixtures were also simulated in the NVT ensemble at a mean pressure $\langle P \rangle$ = 0 and $T$ = 0.80 with $N$ = 1000, 5000, and 10,000 at which the relaxation times of the systems (see later section) are similar.\\

\noindent 
Simulations of Cu$_x$Zr$_{1-x}$ mixtures in the NPT ensemble were performed using the LAMMPS package with compositions of $x$ = 64.5, and 35.5. The Embedded Atom Model (EAM) method of Finnis-Sinclair was applied \cite{mendelev2009} simulating at a pressure $P$ = 0. The system was first melted at $T$ = 2000K and then cooled and equilibrated at $T$ = 1270K which is somewhat above the eutectic temperatures. Three system sizes were simulated $N$ = 1000, 5000, and 10,000. \\

\noindent
\textbf{Identifying Local Structure. --- }
To detect the FCC and HCP crystals, and bicapped square antiprism liquid locally favoured structure, we use the topological cluster classification (TCC) employed previously to identify local structures in the KA mixture \cite{malins2013fara}. That is to say, we carry out a standard Voronoi decomposition and seek structures topologically identical to geometric motifs of particular interest. 

For the BCC crystal, we employ a bond-orientational order (BOO) parameter analysis \cite{steinhardt1983}. For each particle $i$ we define complex order parameters $q^{i}_{lm} \equiv 1/n_{b}\sum_{j=1}^{n_{b}}Y_{lm}(\theta_{ij}, \phi_{ij})$, where  $Y_{lm}$ is the spherical harmonic function with degree $l$ and order $m$, $\theta$ and $\phi$ are the spherical coordinates for the vector $\textbf{r}_{ij} \equiv \textbf{r}_{j} - \textbf{r}_{i}$, and $n_{b}$ is the number of neighbours defined from the 12 nearest neighbors. We use the complex order parameters  to differentiate between solid and liquid particles using the criteria that for at least $7$ nearest-neighbor bonds the scalar product $\textbf{q}_{6}^{i} \cdot \textbf{q}_{6}^{j}/|\textbf{q}_{6}^{i}||\textbf{q}_{6}^{j}|$  should be greater than 0.70 to be classified as a solid particle. $\textbf{q}_{6}^{i}$ is a $(2l+1)$-dimensional complex vector. The identity of each solid particle is then determined \cite{russo2012} using the third-order invariant order parameters

\begin{displaymath}
W_{l}^{i} \equiv \sum_{m_{1},m_{2},m_{3}=0}^{l} \left (\begin{array}{c c c} l & l & l  \\
m_{1} & m_{2} & m_{3}
\end{array}\right)
\frac{Q^{i}_{lm_{1}}Q^{i}_{lm_{2}}Q^{i}_{lm_{3}}}{|\textbf{Q}^{i}_{l}|^{3}},
\end{displaymath}
where the term in the parentheses is the Wigner $3 - j$  symbol and $Q^{i}_{lm} \equiv 1/(n_{b}+1)\sum_{i=1}^{n_{b}+1}q^{i}_{lm}$ is the average bond-orientational order parameter \cite{lechner2008}. BCC particles are identified as all solid particles having $W_6^{i} > 0$. The BOO analysis also provides the identity of FCC and HCP solid particles (using $W_{4}^{i}$; see, e.g., Ref. \cite{russo2012}) but in this case the TCC algorithm is favored. \\

\noindent
\textbf{Relaxation Time Determination. --- }
We determine the relaxation time of the liquid $\tau_\alpha$ from the self-part of the intermediate scattering function $F_{s}(q,t) \equiv\langle\exp[i q \Delta \textbf{r}]\rangle$ of the $A$-particles using the criterion $F_{sA}(q, \tau_\alpha)=0.2$; the length of the wave vector is $q = 7.25$. A system size of $N$ = 1000 is used for these simulations to supress nucleation but has a minor effect on $\tau_\alpha$. In the case where we cannot measure $\tau_\alpha$
directly in simulations due to extremely long simulation timescales
(say at $T$ = 0.395 or 0.400) we extrapolate using a VFT fit 
(see SM for more details). \\

\noindent
\textbf{Determining the Error in the Crystallisation Times. --- } For state points where all runs crystallised, we estimate the error with the standard deviation in the crystallisation times for each run. In the case where not all runs crystallised (say $m$ runs crystallised), we estimate the error from the difference in the value of the crystallisation time determined, $\tau_X(m)$, and that determined had an additional run crystallised $\tau_X(m+1)$. That is to say, the error is $\delta \tau_X = \tau_X(m) - \tau_X(m-1)$.


\end{document}